\documentclass[a4paper,10pt]{article}

\usepackage{amssymb,amsfonts,amsmath,epsfig,float,graphicx,subfigure}

\date{}
\author{
Tobias Verhulst\footnote{Research Assistant of the Research Foundation - Flanders (FWO - Vlaanderen) \newline tobias.verhulst@ua.ac.be}, Ben Anthonis,
and Jan Naudts\footnote {jan.naudts@ua.ac.be} \\
[1ex]
\footnotesize\it
 Departement Fysica, Universiteit Antwerpen,\\
\footnotesize\it
 Groenenborgerlaan 171, 2020 Antwerpen, Belgium}
\title{Analysis of the $N=4$ Hubbard ring using counting operators}

\newtheorem{proposition}{Theorem}{}

\def\beginproof{\par\noindent{\bf Proof}\par}
\def\endproof{\par\strut\hfill$\square$\par\vskip 0.5cm}

\begin{document}
\maketitle
\begin{abstract}
We prove three theorems about the use of a counting operator to study the spectrum of model Hamiltonians. We analytically calculate the eigenvalues of the Hubbard ring with four lattice positions and apply our theorems to describe the observed level crossings.
\end{abstract}

\section{Introduction}
The Hubbard model is a simplified lattice model used to describe electrons with short-range interactions \cite{HJ63}. Its main application is the description of a variety of phenomena in solids \cite{T98,BDZ08}. A lot of work has been done on finding exact eigenvalues for the Hubbard model and for similar models. In most cases only the one dimensional models can be solved analytically. We refer to the one dimensional Hubbard model with periodic boundary conditions as the Hubbard ring. Most analytic results for the Hubbard ring are found using the Bethe ansatz \cite{LSM61,LW68,EK94,DEGKKK00,HHK08,JSA08}. However, also other methods have been used to analyse these models \cite{T98,S71,HL71,SR02,KH08}.

One of the interesting aspects of the Hubbard ring is that one observes crossings between the energy levels which seemingly violate the Wigner-von Neumann noncrossing rule \cite{HL71,YAS02}. One can use the symmetries of the Hamiltonian to explain such violations \cite{YAS02,OWU08,DFM01}.
Here, we give an explanation which does not rely on the integrability of $H$
but rather on that of some reference Hamiltonian $H_{\rm ref}$. The latter was introduced in \cite{NV08}.
This approach is more general in the sense that $H$ does not need to be
integrable and $H_{\rm ref}$ can have a much simpler spectrum than $H$.

We elaborate the method of \cite{NV08} by introducing a hierarchy of equations. Although we will only use this hierarchy to find solutions for the $N=4$ Hubbard ring it should be noted that such a hierarchy always exists.

Overview of the paper. In Section 2 the Hubbard Hamiltonian is defined and some notations are introduced. In Section 3 the hierarchy is introduced and the general theorems are proven. Section 4 contains the analytical eigenvalues for the $N=4$ Hubbard ring. We use our theorems to classify these eigenvalues. Section 5 gives an overview of the results and conclusions.

\section{Counting operators for the $N=4$ Hubbard ring}
The Hamiltonian of the Hubbard ring is \cite{HJ63,T98}
\begin{equation}
H=-\sum_{i,j}t_{ij}\sum_{\sigma=\uparrow, \downarrow}b^\dagger_{i,\sigma}b_{j,\sigma}+\alpha\sum_kn_{k,\uparrow}n_{k,\downarrow}
\label{hamiltonian}
\end{equation}
We consider this model on a lattice with four positions and periodic boundary conditions. Furthermore we assume that only nearest neighbour hopping is possible and that the band is half filled. This means there are two electrons with spin up and two with spin down and the coefficients $t_{ij}$ satisfy $t_{ij}=\delta_{i,j+1}+\delta_{i,j-1}$. Thus the only parameter on which the eigenvalues of (\ref{hamiltonian}) depend is $\alpha$.

In \cite{NV08}, the counting operator, here denoted $M_0$, is defined as
\begin{equation}
M_0=\sum_kn_{k,\uparrow}n_{k,\downarrow}.
\end{equation} It counts the number of sites with double occupation.
However, the choice of counting operator is never unique. It turns out to be advantageous also
to consider additional counting operators $M_1$ and $M_2$, which differ from $M_0$ by adding,
respectively, subtracting 1, whenever a doubly occupied site is flanked by to singly occupied
sites. In formulas,
\begin{eqnarray}
M_1&=&M_0+\mathcal P_A
\label{M1operator}\\
M_2&=&M_0-\mathcal P_A
\label{M2operator}
\end{eqnarray}
where
% $A=\left\{\varphi:\sum_kn_{k,\uparrow}n_{k,\downarrow}\varphi=\varphi,
%\sum_{k,\sigma,\sigma'}n_{k,\sigma}n_{k+1,\sigma'}\varphi=4\varphi\right\}$ and 
$\mathcal P_A$ is the orthogonal projection onto the space $A$
spanned by the configurations where two spins on the same site
are surrounded by sites with a single occupancy.
It can easily be verified that $[[[H,M_i],M_i],M_i]=[H,M_i]$ for $i=0,1,2$. Then \cite{NV08} there exist operators $H_{\rm ref, i}$ and $R_i$ such that
\begin{equation}
H=H_{\rm ref, i}+R_i+R_i^\dagger
\label{decomposition}
\end{equation}
with $R_i$ and $H_{\rm ref, i}$ satisfying
\begin{eqnarray}
\left[H_{\rm ref, i},M_i\right]&=&0\nonumber\\
\left[R_i,M_i\right]&=& R_i\label{commutation}\\
H_{\rm ref, i}^\dagger&=&H_{\rm ref, i}\nonumber
\end{eqnarray}
For the choices (\ref{M1operator}) and (\ref{M2operator}), the corresponding reference Hamiltonians $H_{\rm ref, i}$ coincide. The extra index $i$ will be omitted in the following,
this is, $H_{\rm ref}\equiv H_{\rm ref,1}=H_{\rm ref,2}$.
However, $R_1$ does not equal $R_2$. The method to construct explicit expressions for $H_{\rm ref}$ en $R_i$ can be found in \cite{NV08}. It is shown in \cite{NV08} that, in general, the eigenvalues of $M_i$ are of the form $\mu_m=\nu_n+m$ for some constants $\nu_n$. In the $N=4$ Hubbard ring with two spin up and two spin down electrons the eigenvalues of the $M_i$ are $0$, $1$ and $2$.

\section{The hierarchy}
To simplify the notation we will in this section omit the index $i$. All results hold for any of the $M_i$. In this section we derive a hierarchy of equations which is equivalent with the eigenvalue equation. This hierarchy will later be used to analyse the spectrum of the $N=4$ Hubbard ring. However, the theorems in this section are much more general. The only condition for these theorems to hold is that the Hamiltonian can be written as (\ref{decomposition}), satisfying (\ref{commutation}). To simplify the calculations we will also assume that the spectrum of $M$ is of the form $\mu_m=\nu+m$ for some $\nu$, as is the case for the Hubbard ring. However, the theorems can be generalized for models where this is not the case.

An eigenvector $\Psi$ of $H$ can be decomposed as
\begin{equation}
	\Psi=\sum_{m=1}^{m_{\rm max}}\psi_m
	\label{eigenvector}
\end{equation}
where $M\psi_m=(\nu+m_0+m-1)\psi_m$. Here, $\psi_1$ and $\psi_{m_{\rm max}}$ are both non zero, $m_{\rm max}$ can be finite or infinite and $m_0$ can be both zero or non zero. Without loss of generality we will prove our theorems for $m_0=0$. Using this decomposition of the eigenvector the eigenvalue equation $H\Psi=\varepsilon\Psi$ can also be decomposed into the following hierarchy of equations
\begin{equation}
	\left(H_{\rm ref}-\varepsilon\right)\psi_m+R\psi_{m+1}+R^\dagger\psi_{m-1}=0
	\label{eigenvalue}
\end{equation}
If $m_{\rm max}=1$, then (\ref{eigenvalue}) simplifies to
\begin{equation}
	H_{\rm ref}\psi_1=\varepsilon\psi_1
	\label{simple}
\end{equation}
Now assume that $m_{\rm max}\neq1$. The first equation of the hierarchy then becomes
\begin{equation}
	(H_{\rm ref}-\varepsilon)\psi_1+R\psi_2=0
	\label{lowest}
\end{equation}
If $m_{\rm max}$ is finite there is also a last equation in the hierarchy which reads
\begin{equation}
	(H_{\rm ref}-\varepsilon)\psi_{m_{\rm max}}+R^\dagger\psi_{m_{\rm max}-1}=0
	\label{highest}
\end{equation}
From (\ref{commutation}) it is clear that $M$ and $RR^\dagger$ commute.
Thus it is possible to find simultaneous eigenstates for $M$ and $RR^\dagger$. The same is true for $M$ and $H_{\rm ref}$. However $H_{\rm ref}$ does not commute with $RR^\dagger$. While this means there is no basis in which both operators are diagonal it is still possible that there are some simultaneous eigenstates. If such states exist they can be used to construct special eigenstates of $H$ according to the following Theorem.
\begin{proposition}
Let $\Phi$ be an eigenstate of $M$, $RR^\dagger$, $(R^\dagger)^2$ and $R$ with eigenvalues $\mu$, $\xi$, $0$ and $0$. Assume the states $\Phi$ and $R^\dagger\Phi$ are both eigenstates of $H_{\rm ref}$ with eigenvalues $\mu+k$ and $\mu+\alpha+k+\Delta k$. Then $\Psi=(1+hR^\dagger)\Phi$ with $h=\frac{\alpha+\Delta k}{2\xi}\pm\frac{1}{2\xi}\sqrt{(\alpha+\Delta k)^2+4\xi}$ are eigenstates of $H$ which have $m_{\rm max}=2$ and eigenvalue $\varepsilon=\mu+k+h\xi$.
\end{proposition}
\beginproof
	Calculating $H\Psi$ gives
	\begin{eqnarray}
		H\Psi&=&H(1+hR^\dagger)\Phi\nonumber\\
		&=&\left(H_{\rm ref}+R^\dagger+hH_{\rm ref}R^\dagger+hRR^\dagger\right)\Phi\nonumber\\
		&=&(H_{\rm ref}+hRR^\dagger)\Phi+(1+hH_{\rm ref})R^\dagger\Phi\nonumber\\
		&=&(\mu+k+h\xi)\Phi+\left(1+h(\mu+\alpha+k+\Delta k)\right)R^\dagger\Phi\label{heq}
	\end{eqnarray}
	From this it follows that $\Psi$ is an eigenstate of $H$ if
	$$
	h(\mu+k+h\xi)=h(\mu+\alpha+k+\Delta k)+1
	$$
	This condition is satisfied if $h=\frac{\alpha+\Delta k}{2\xi}\pm\frac{1}{2\xi}\sqrt{(\alpha+\Delta k)^2+4\xi}$. From (\ref{heq}) follows that $\varepsilon=\mu+k+h\xi$. Because $R\Phi=(R^\dagger)^2\Phi=0$ it is straightforward to prove that $m_{\rm max}=2$.
\endproof
In the case of the $N=4$ Hubbard ring with two spin up electrons and two spin down electrons both $k$ and $\Delta k$ equal zero. However, if an electron is added to (or removed from) the model there are states of this form for which $\Delta k$ does not vanish. A similar construction can be done for $m_{\rm max}>2$ states. However, the constant $h$ then becomes a solution of higher order equations which means there are not always analytical expressions for the eigenvalues. For the $N=4$ Hubbard ring the highest order equation that occurs in these constructions is a cubic equation.

The following theorem about the $m_{\rm max}=2$ case can now be proven (a more general form of this theorem, not using the hierarchy, can be found in \cite{NV08}).
\begin{proposition}
If $\Psi$ is an eigenstate of $H$ for which $m_{\rm max}=2$, then $\Phi=e^{-zM}\Psi$ is also an eigenstate of $H$ for some $z\in\mathbb C$, with a different eigenvalue.
\end{proposition}
\beginproof
	Because $\Psi$ has $m_{\rm max}=2$ it can be written as $\Psi=\psi_1+\psi_2$.  Clearly $\Phi$ can then be decomposed as $\Phi=\phi_1+\phi_2=e^{-zM}\psi_1+e^{-zM}\psi_2$ with	$R\phi_1=R^\dagger\phi_2=0$. From Theorem 1 then follows that also $(H_{\rm ref}-\alpha M)\phi_1=(H_{\rm ref}-\alpha M)\phi_2=0$. Therefore
	\begin{eqnarray*}
		H\Phi&=&\alpha M\Phi+R^\dagger\phi_1+R\phi_2\\
		&=&\nu\alpha\phi_1+(\nu\alpha+\alpha)\phi_2+e^{-z(M-1)}R^\dagger\psi_1+e^{-z(M+1)}R\psi_2\\\
		&=&\nu\alpha\phi_1+(\nu\alpha+\alpha)\phi_2-e^{z}(\nu\alpha+\alpha-\mu)\phi_2-e^{-z}(\nu\alpha-\mu)\phi_1
	\end{eqnarray*}
	Hence, $\Phi$ is an eigenstate of $H$ iff
	\begin{equation}
		(\mu-\nu\alpha)e^{-z}=\alpha+(\mu-\nu\alpha-\alpha)e^{z}
		\label{condition}
	\end{equation}
	Thus
	\begin{equation}
		e^{z}=-\frac{\nu\alpha-\mu}{\nu\alpha-\mu+\alpha}
	\end{equation}
	from which $z$ can be calculated.
\endproof
The eigenvalues $\varepsilon_1$ and $\varepsilon_2$ of two states which are connected in the sense of Theorem 2 satisfy
\begin{equation}
	\left\{\begin{array}{rcl}
	\varepsilon_1+\varepsilon_2&=&\alpha\\
	\varepsilon_1-\varepsilon_2&=&\sqrt{\alpha^2+4\xi}
	\end{array}
	\right.
	\label{relation}
\end{equation}
with $\xi$ defined as in Theorem 1.

In the formulation of the third Theorem we use the following notion.
The hierarchical decomposition (\ref {eigenvector}) of the eigenvector $\Psi$
is said to be {\sl minimal} if none of the sums
\begin{eqnarray*}
 \sum_{m=1}^{m'}\psi_m
\end{eqnarray*}
with $m'<m_{\rm max}$ is an eigenvector of $H$.

\begin{proposition}
If $\Phi$ and $\Psi$ are eigenstates of $H$ with the same eigenvalue  $\varepsilon$. Assume that the hierarchical decomposition of $\Psi$
is minimal and that there exists a function $f$ such that $\Phi=f(M)\Psi$. Then there exists a constant $\mu$ such that $\Phi=f(\mu)\Psi$.
\end{proposition}

\beginproof
In the case that $m_{\rm max}=1$ nothing has to be proved.
Hence assume that $m_{\rm max}>1$.

Writing (\ref{lowest}) for both $\Psi$ and $\Phi=f_\alpha(M)\Psi$ gives (to simplify the notation we assume $\nu=0$)
\begin{eqnarray*}
(H_{\rm ref}-\varepsilon)\psi_1+R\psi_2&=&0,\\
f(\mu_1)(H_{\rm ref}-\varepsilon)\psi_1+f(\mu_2)R\psi_2&=&0
\end{eqnarray*}
where $\mu_1,\mu_2,\cdots$ are the eigenvalues of $M$.
From this pair of equations follows
\begin{eqnarray*}
\left(f(\mu_1)-f(\mu_2)\right)R\psi_2=0.
\end{eqnarray*}
Note that $R\psi_2=0$ implies that $\psi_1$ is an eigenvector of $H$.
But this is not possible because of the assumption that the
hierarchical decomposition of $\Psi$ is minimal. Hence, one concludes that
$f(\mu_1)=f(\mu_2)$.

The proof now continues by induction.
Assume that $f(\mu_1)=f(\mu_2)=\cdots=f(\mu_{m'})$ has been proved for some $m'<m_{\rm max}$ and let us show that this assumption also holds for $m'$
increased by 1. From (\ref {eigenvalue}) follows
\begin{eqnarray*}
(H_{\rm ref}-\varepsilon)\psi_{m'}+R\psi_{m'+1}+R^\dagger\psi_{m'-1}&=&0,\\
f(\mu_{m'})(H_{\rm ref}-\varepsilon)\psi_m
+f(\mu_{m'+1})R\psi_{m'+1}+f(\mu_{m'-1})R^\dagger\psi_{m'-1}&=&0.
\end{eqnarray*}
Using the assumption that $f(\mu_{m'-1})=f(\mu_{m'})$ these equations
imply that
\begin{eqnarray*}
\left(f(\mu_{m'+1})-f(\mu_{m'})\right)R\psi_{m'+1}=0.
\end{eqnarray*}
Now, $R\psi_{m'+1}=0$ implies that $\sum_{m=1}^{m'}\psi_m$ is an eigenvector
of $H$. This contradicts the assumption that the hierarchical decomposition
of $\Psi$ is minimal. Hence one concludes that $f(\mu_{m'+1})=f(\mu_{m'})$.

By induction one concludes that all $f(\mu_m)$ are equal. This finishes
the proof of the Theorem.
\endproof

The Theorem implies that two eigenvectors $\Phi$ and $\Psi$ of $H$,
belonging to the same multiplet and connected by the operator $f(M)$,
cannot have the same eigenvalue.
This property is expected because of the Wigner-von Neumann noncrossing rule. Because $M$ commutes with $H_{\rm ref}$ there will be some degeneracies in the spectrum of $H_{\rm ref}$ which disappear when a perturbation $\lambda(R+R^\dagger)$ is added to $H_{\rm ref}$. The noncrossing rule states that those eigenvalues which are degenerated for $\lambda=0$ do not cross each other for any other value of $\lambda$. Now, $\lambda$ equals one in the Hamiltonian $H$. Hence, the two eigenvalues of $H$, which for $\lambda=0$ are degenerate, must not coincide for $\lambda=1$. Note however that a multiplet of $H$ must not necessarily originate from
a degenerate multiplet of $H_{\rm ref}$. Hence, the result of the Theorem is not an immediate consequence
of the Wigner-von Neumann noncrossing rule.

\section{The spectrum of $H$}
Analytical expressions for the eigenvalues as functions of the parameter $\alpha$
have been obtained in \cite {SR02}. They can also be found using the symbolic software package Maple. The eigenvalues are given in Table 1 together with their corresponding degeneracies.
\begin{center}
\begin{table}[!h!t]
\newcommand\T{\rule{0pt}{2.6ex}}
\newcommand\B{\rule[-1.2ex]{0pt}{0pt}}
\caption{The eigenvalues of Hamiltonian (\ref{hamiltonian}). The functions $G_\pm(\alpha)$ and $F_\pm(\alpha)$ are defined by
$G_\pm(\alpha)=\mp\alpha^3\pm36\alpha+6\sqrt{-6\alpha^4-156\alpha^2-3072}$ and  $F_\pm(\alpha)=\pm108\alpha+\sqrt{-3\alpha^6-144\alpha^4-1008\alpha^2-12288}$. These eigenvalues are in agreement with the results of \cite{HL71}.}
\begin{center}\begin{tabular}[l]{||l|c|l||}
	\hline\hline
	class&eigenvalue\T\B&degeneracy\\\hline\hline
	$\langle1\rangle$&0&1\T\\
	$\langle2\rangle$&$\alpha$&6\\
	$\langle3\rangle$&$2\alpha$&1\B\\\hline
	\T $\langle1,2\rangle$&$\frac12\alpha+\frac12\sqrt{\alpha^2+16}$&2\\
	&$\frac12\alpha-\frac12\sqrt{\alpha^2+16}$&2\\
	$\langle2,2\rangle$&$\alpha+2$&4\\
	&$\alpha-2$&4\\
	$\langle2,3\rangle$&$\frac32\alpha+\frac32\sqrt{\alpha^2+16}$&2\\
	&$\frac32\alpha-\frac32\sqrt{\alpha^2+16}$&2\B\\\hline
	$\langle1,2,3\rangle_a$&$\frac13F_+(\alpha)^\frac13+(\alpha^2+16)F_+(\alpha)^{-\frac13}+\alpha$\T&1\\
	&$-\frac13e^{\frac{-i\pi}{3}}F_+(\alpha)^\frac13-e^{\frac{i\pi}{3}}(\alpha^2+16)F_+(\alpha)^{-\frac13}+\alpha$&1\\
	&$-\frac13e^{\frac{i\pi}{3}}F_+(\alpha)^\frac13-e^{\frac{-i\pi}{3}}(\alpha^2+16)F_+(\alpha)^{-\frac13}-\alpha$&1\\
	$\langle1,2,3\rangle_b$&$\frac13F_-(\alpha)^\frac13+(\alpha^2+16)F_-(\alpha)^{-\frac13}+\alpha$&1\\
	&$-\frac13e^{\frac{-i\pi}{3}}F_-(\alpha)^\frac13-e^{\frac{i\pi}{3}}(\alpha^2+16)F_-(\alpha)^{-\frac13}+\alpha$&1\\
	&$-\frac13e^{\frac{i\pi}{3}}F_-(\alpha)^\frac13-e^{\frac{-i\pi}{3}}(\alpha^2+16)F_-(\alpha)^{-\frac13}-\alpha$&1\\
	$\langle1,2,2\rangle$&$\frac13\left(G_+(\alpha)^\frac13+(\alpha^2+48)G_+(\alpha)^{-\frac13}+2\alpha\right)$&1\\
	&$-\frac13\left(e^{\frac{-i\pi}{3}}G_+(\alpha)^\frac13+e^{\frac{i\pi}{3}}(\alpha^2+48)G_+(\alpha)^{-\frac13}-2\alpha\right)$&1\\
	&$-\frac13\left(e^{\frac{i\pi}{3}}G_+(\alpha)^\frac13+e^{\frac{-i\pi}{3}}(\alpha^2+48)G_+(\alpha)^{-\frac13}+2\alpha\right)$&1\\
	$\langle2,2,3\rangle$&$\frac13\left(G_-(\alpha)^\frac13+(\alpha^2+48)G_-(\alpha)^{-\frac13}+4\alpha\right)$&1\\
	&$-\frac13\left(e^{\frac{-i\pi}{3}}G_-(\alpha)^\frac13+e^{\frac{i\pi}{3}}(\alpha^2+48)G_-(\alpha)^{-\frac13}-4\alpha\right)$&1\\
	&$-\frac13\left(e^{\frac{i\pi}{3}}G_-(\alpha)^\frac13+e^{\frac{-i\pi}{3}}(\alpha^2+48)G_-(\alpha)^{-\frac13}+4\alpha\right)$&1\B\\\hline\hline
\end{tabular}\end{center}
\end{table}
\end{center}
Figure 1 shows the eigenvalues of the $N=4$ Hubbard ring as a function of $\alpha$. It can be seen that there are several crossings between energy levels. The crossings at $\alpha=0$ are explained by the higher symmetry of the Hamiltonian without interaction. However, there are also some crossings for $\alpha\neq0$. This is similar to the observations of \cite{HL71}.

\begin{figure}[!h!t]
	\centering
	\includegraphics[width=.9\textwidth,height=8.5cm]{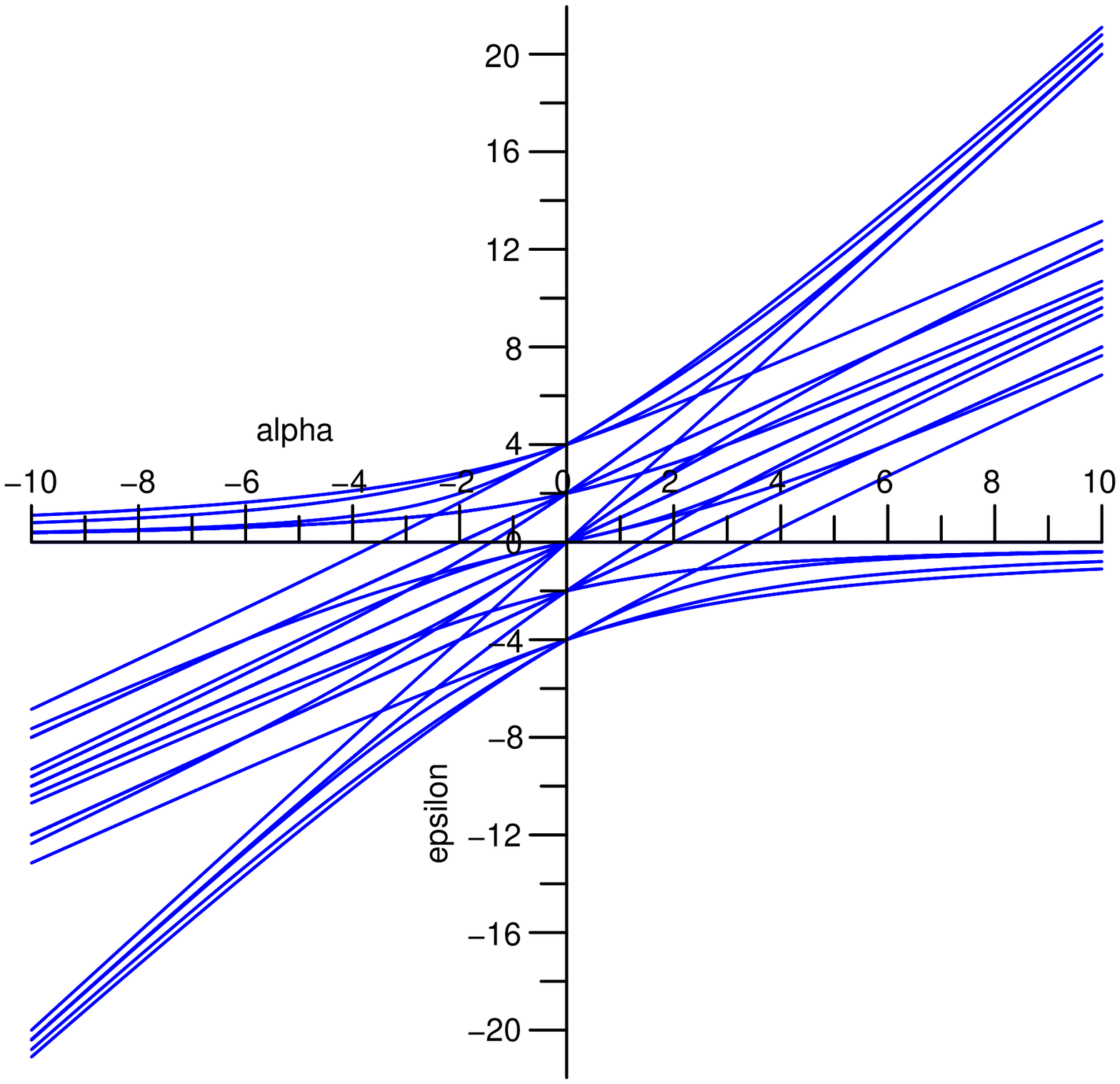}
	\caption{The complete spectrum of the $N=4$ Hubbard ring. Both $\alpha$ and $\varepsilon$ are in arbitrary units. Notice the crossings between the levels. Most crossings occur for $\alpha=0$ but there are also some crossings for other values of $\alpha$.}
\end{figure}

Using (\ref{eigenvector}) the Hilbert space $\cal H$ of eigenfunctions of the Hamiltonian (\ref{hamiltonian}) can be decomposed into subspaces $\mathcal{H}_{n,i}$, where each $M_i$ gives a slightly different decomposition. $R_i$ and $R_i^\dagger$ act as creation and annihilation operators between the subspaces mapping $\mathcal{H}_{n,i}$ to $\mathcal{H}_{n\pm1,i}$. In the case of the $N=4$ Hubbard ring, the $M_i$ have three eigenvalues (0, 1 and 2) so there are three different subspaces. The eigenvalues are then classified by indicating the subspaces $\mathcal H_{n,0}$ on which  which the eigenstates have a non-vanishing orthogonal projection.

The first three eigenvalues in Table 1 correspond to the states which fall in one of the $\mathcal H_{n,0}$. The corresponding eigenstates can easily be constructed. For example the eigenvalue $2\alpha$ corresponds to
\begin{equation}
	\Psi=\varphi_{12}-\varphi_{13}+\varphi_{14}+\varphi_{23}-\varphi_{24}+\varphi_{34}\nonumber
\end{equation}
with the notation $\varphi_{ij}=b^\dagger_{i,\uparrow}b^\dagger_{j,\uparrow}b^\dagger_{i,\downarrow}b^\dagger_{j,\downarrow}|0\rangle$. This eigenvector falls in $\mathcal{H}_{3,0}$, the eigenvector for which $\varepsilon=0$ falls in $\mathcal{H}_{1,0}$ and the others in $\mathcal{H}_{2,0}$.

The next six eigenvalues, the $\langle1,2\rangle$, $\langle2,2\rangle$ and $\langle2,3\rangle$ pairs, correspond to the states that for some $n$ fall within $\mathcal{H}_{n,0}\oplus\mathcal{H}_{n+1,0}$. The three pairs each form a doublet of the form discussed in Theorem 2. Note that there are no eigenstates in $\mathcal{H}_{1,0}\oplus\mathcal{H}_{3,0}$. The reason for this is that if $\psi_2=0$, then $R^\dagger\psi_1=0$. Thus the hierarchy reduces to the single equation (\ref{simple}) and the eigenvalue falls into one of the first three classes.

The remaining eigenvalues correspond to triplet states. The only cases where $M_1$ and $M_2$ give different results is for the triplets $\langle1,2,2\rangle$ and $\langle2,2,3\rangle$. Using the hierarchy derived from $M_1$ one cannot construct the states from the $\langle2,2,3\rangle$ triplet and from $M_2$ one cannot construct the $\langle1,2,2\rangle$ triplet.

\begin{figure}[!h!t]
	\centering
	\subfigure[]{\includegraphics[width=.45\textwidth]{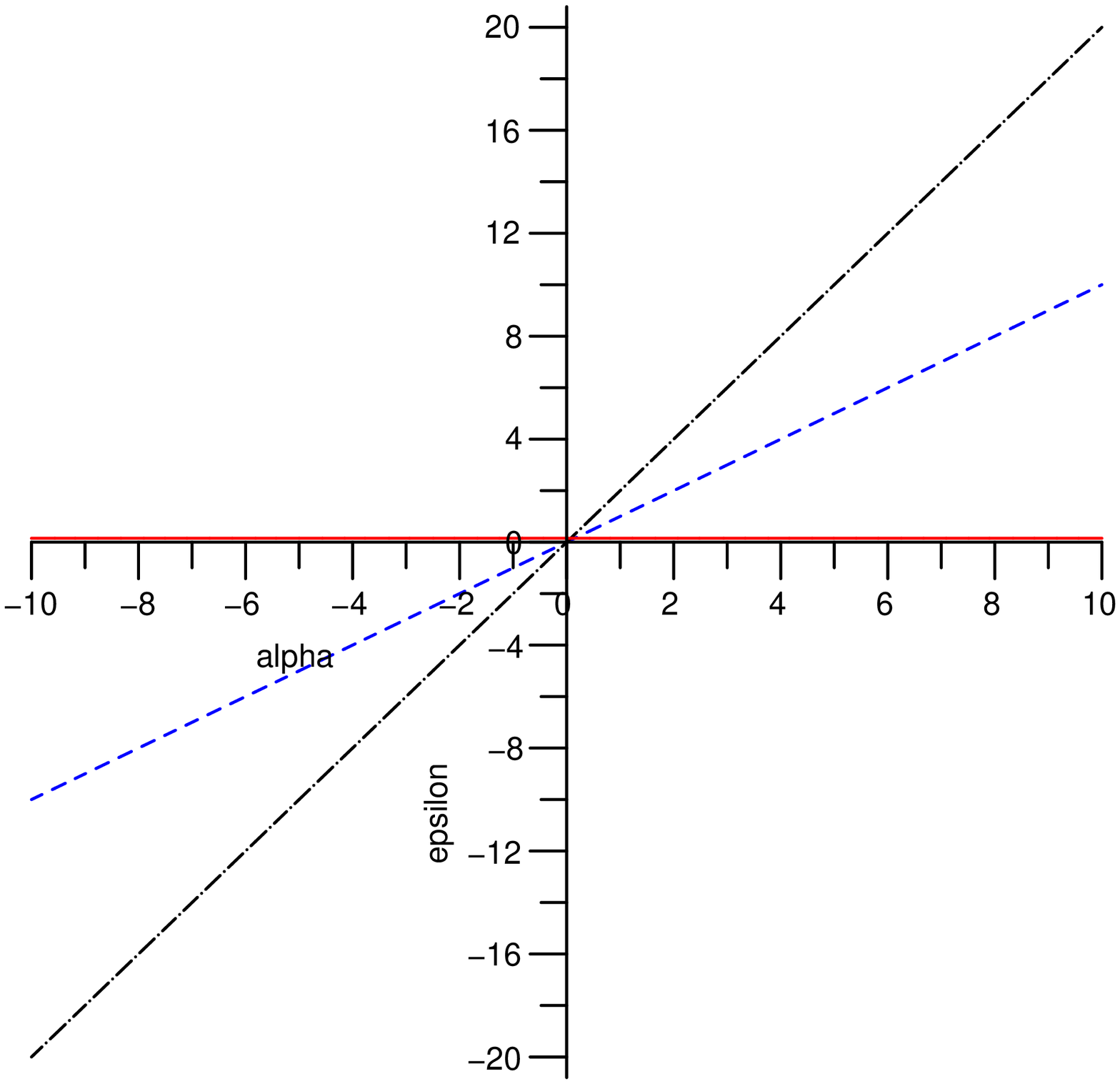}}
	\hspace{1cm}
	\subfigure[]{\includegraphics[width=.45\textwidth]{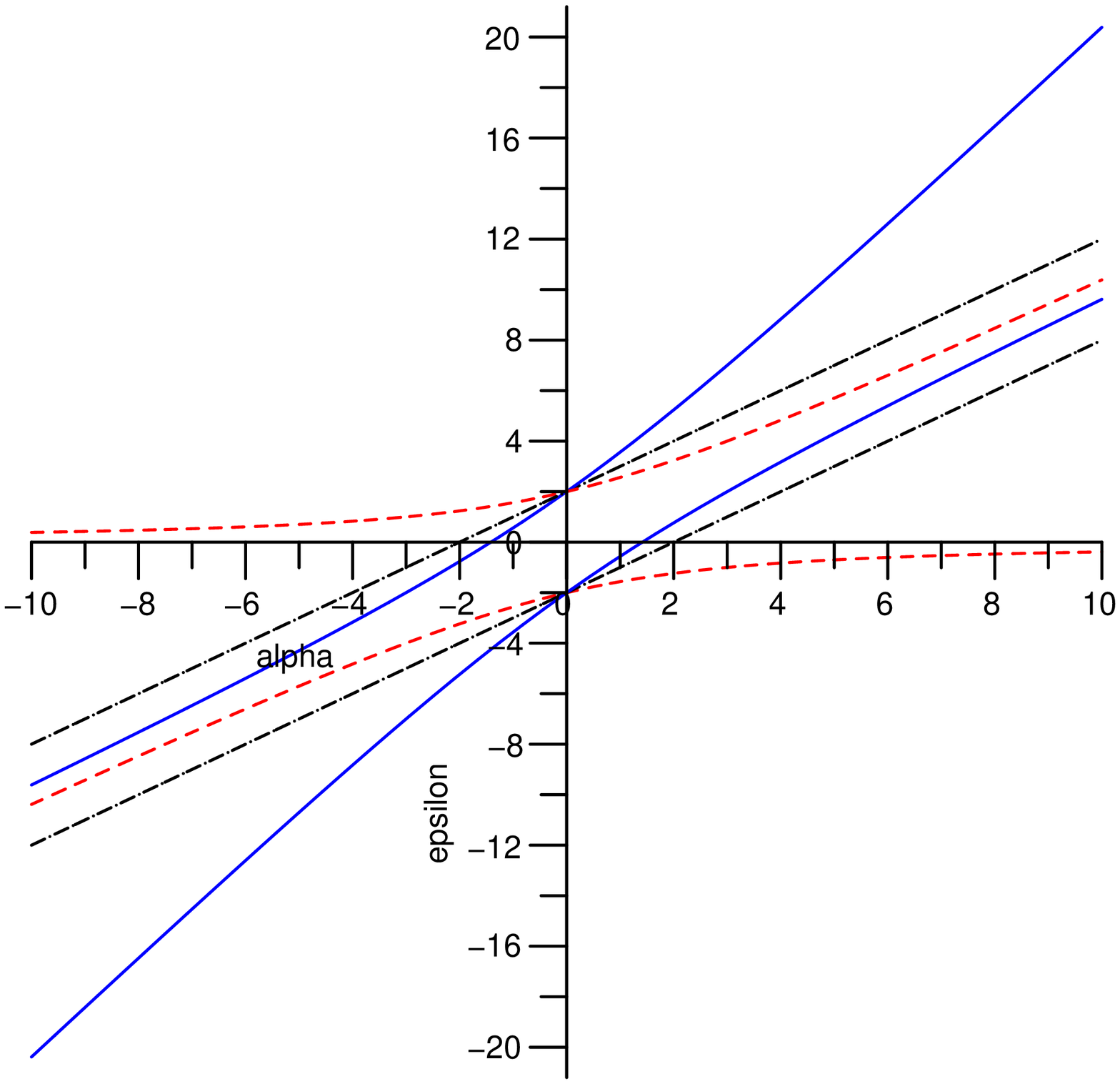}}
	\vspace{1cm}
	\subfigure[]{\includegraphics[width=.45\textwidth]{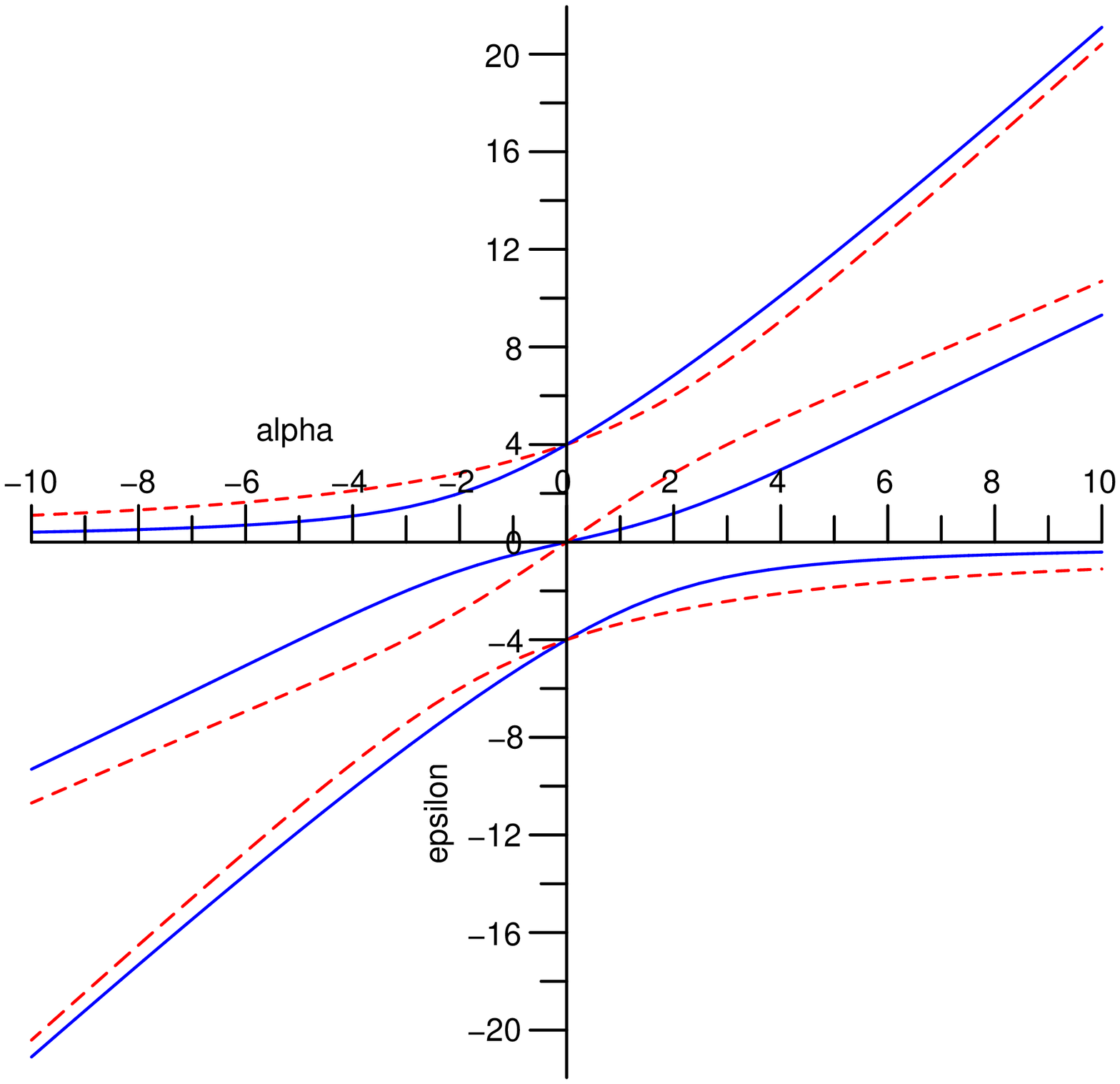}}
	\hspace{1cm}
	\subfigure[]{\includegraphics[width=.45\textwidth]{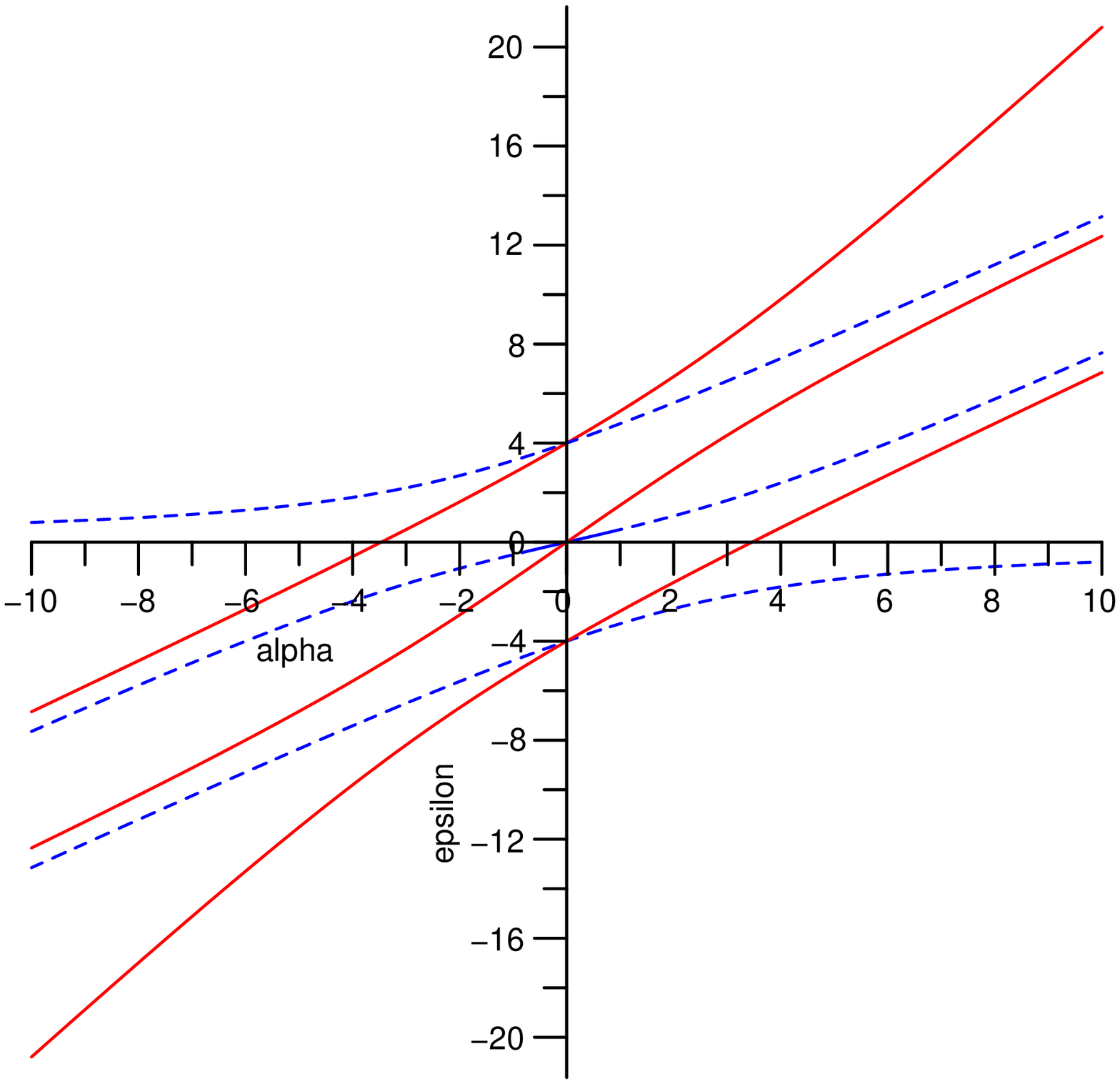}}
	\caption{The eigenvalues of the $N=4$ Hubbard ring with both $\alpha$ and $\varepsilon$ in arbitrary units. (a) shows the singlets. The solid red line corresponds to $\langle1\rangle$, the dashed blue line to $\langle2\rangle$ and the black dots to $\langle3\rangle$. (b) shows the doublet states. The pair shown in red dashes is the $\langle1,2\rangle$ doublet (which falls in $\mathcal{H}_1\oplus\mathcal{H}_2$), the black dotted lines show the $\langle2,2\rangle$ pair and the solid blue pair is $\langle2,3\rangle$. In (c) $\langle1,2,3\rangle_a$ is shown as blue solid lines and $\langle1,2,3\rangle_b$ as red dashes while in (d) $\langle1,2,2\rangle$ is shown in blue dashes and $\langle2,2,3\rangle$ in red solid. Notice that within one doublet or triplet there are no crossings between different levels.}
\end{figure}

In Figure 2b it can be seen that within each doublet there are no crossings between energy levels. In Figure 2c and 2d it can be seen that this is also true for the triplets. This was predicted by Theorem 3. In \cite{NV08} it is shown that for some models Theorem 2 can be generalized in the sense that from one eigenstate $\Phi$ others can be constructed by defining a function $f(M)$ such that $\Psi=f(M)\Phi$ is also an eigenstate. Theorem 2 is a special case where $f(x)=e^{zx}$. In \cite{NV08} also an example model was given for which there exists triplets of states that are connected to each other by such functions and thus the corresponding eigenvalues do not cross each other. This is also the case here for the triplets in the $N=4$ Hubbard model although we do not know the explicit expressions for these functions.

\section{Conclusions}
Using the decomposition of the Hamiltonian as described in \cite{NV08}, we derive a hierarchy of equations which is equivalent to the eigenvalue equation. This hierarchy simplifies the calculation of the eigenstates. Its most important feature, however, is that it gives a natural classification of the eigenstates in terms of multiplets.

The first two of our theorems give a method for constructing doublet states which are orthogonal to all but two of the subspaces of $\mathcal H$. In Theorem 1 a method is given to construct the states of such a doublet and Theorem 2 gives an expression for the function connecting those states. A construction similar to the one of Theorem 1 can also be done for other states, although for the most general case the calculations may need to be done numerically. However in the $N=4$ Hubbard ring there exist only singlets, doublets and triplets which we could all identify. Finally, Theorem 3 states that two eigenvalues belonging to the same multiplet never cross (for the doublet states this can also easily be seen from (\ref{relation})).

In conclusion we use an alternative way to define multiplets without using the symmetry of the Hamiltonian. Instead we use operators commuting with a reference Hamiltonian $H_{\rm ref}$. 
%This allows us to analyse non-integrable Hamiltonians.
Future work also needs to be done on generalizing our theorems to triplets and more general multiplets.

\section*{Acknowledgments}
We are grateful to Prof. J. Richter for pointing out \cite {SR02} to us.

\end{document}